# Explaining the emergence of land-use frontiers

**Authors:**
Patrick Meyfroidt[1,2], Dilini Abeygunawardane[1,3], Matthias Baumann[4], Adia Bey[1], Ana Buchadas[4,5], Cristina Chiarella[1,6], Victoria Junquera[7,8], Angela Kronenburg García[1,2], Tobias Kuemmerle[4,5], Yann le Polain de Waroux[9,10], Eduardo Oliveira[11], Michelle Picoli[1,12], Siyu Qin[13], Virginia Rodriguez García[1], Philippe Rufin[1,4]

[1] Earth and Life Institute, UCLouvain, 1348 Louvain-la-Neuve, Belgium
[2] F.R.S. - FNRS, 1000 Brussels, Belgium
[3] Leibniz Institute of Agricultural Development in Transition Economies (IAMO), Theodor-Lieser-Str. 2, 06120, Halle, Germany
[4] Geography Department, Humboldt-University Berlin, Unter den Linden 6, 10099 Berlin, Germany
[5] Integrated Research Institute on Transformations of Human-Environment Systems (IRI THESys), Humboldt-University Berlin, Unter den Linden 6, 10099 Berlin
[6] International Fund for Agricultural Development - IFAD, 00142 Rome, Italy
[7] High Meadows Environmental Institute, Princeton University, Princeton, USA
[8] Dpt. Ecology and Evolutionary Biology, Princeton University, Princeton, USA
[9] Department of Geography, McGill University, Montreal, QC, Canada
[10] Institute for the Study of International Development (ISID), McGill University, Montreal, QC, Canada
[11] Thomas More University of Applied Sciences, 2800 Mechelen, Belgium
[12] WeForest, Cantersteen 47, 1000 Brussels, Belgium
[13] The Nature Conservancy, 10117 Berlin, Germany.

**ORCID:**
Patrick Meyfroidt: 0000-0002-1047-9794
Dilini Abeygunawardane: 0000-0002-5060-4211
Matthias Baumann: 0000-0003-2375-3622
Adia Bey: 0000-0002-2334-3458
Ana Buchadas: 0000-0003-4219-108X
Cristina Chiarella: 0000-0003-2949-1897
Victoria Junquera: 0000-0003-0402-3659
Angela Kronenburg García: 0000-0003-4547-9468
Tobias Kuemmerle: 0000-0002-9775-142X
Yann le Polain de Waroux: 0000-0001-9887-7270
Eduardo Oliveira: 0000-0001-8838-2493
Michelle Picoli: 0000-0001-9855-2046
Siyu Qin: 0000-0001-6466-7400
Virginia Rodriguez García: 0000-0002-2769-0369
Philippe Rufin: 0000-0001-8919-1058

**Acknowledgments:**
This work was supported by the European Research Council (ERC) under the European Union's Horizon 2020 research and innovation program (Grant Agreement No. 677140 MIDLAND and Grant Agreement No. 101001239 SYSTEMSHIFT), and the F.R.S.-FNRS under grant no. T.0154.21, grant no. 1.B422.24. and grant no. 1.B.264.22F.

**CRediT author statement:** PM: Conceptualization; Funding acquisition; Project administration; Supervision; Writing – original draft. TK: Funding acquisition. All authors: Investigation; Writing – review & editing.

# Explaining the emergence of land-use frontiers

**Abstract**

Land use expansion is linked to major sustainability concerns including climate change, food security and biodiversity loss. This expansion is largely concentrated in so-called “frontiers”, defined here as places experiencing marked transformations due to rapid resource exploitation. Understanding the mechanisms shaping these frontiers is crucial for sustainability. Previous work focused mainly on explaining how active frontiers advance, in particular into tropical forests. Comparatively, our understanding of how frontiers emerge in territories considered marginal in terms of agricultural productivity and global market integration remains weak. We synthesize conceptual tools explaining resource and land-use frontiers, including theories of land rent and agglomeration economies, of frontiers as successive waves, spaces of territorialization, friction, and opportunities, anticipation and expectation. We then propose a new theory of frontier emergence, which identifies exogenous pushes, legacies of past waves, and actors’ anticipations as key mechanisms by which frontiers emerge. Processes of abnormal rent creation and capture and the built-up of agglomeration economies then constitute key mechanisms sustaining active frontiers. Finally, we discuss five implications for the governance of frontiers for sustainability. Our theory focuses on agriculture and deforestation frontiers in the tropics, but can be inspirational for other frontier processes including for extractive resources, such as minerals.

# 1. Introduction

Land-use change is key to many sustainability challenges. Conversion of natural ecosystems and conversion from extensive to more intensive land uses are among the main drivers of global environmental change, impacting carbon, biodiversity, water fluxes, livelihoods and many other ecosystem services and nature's contributions to people (1,2). Much of this conversion is not distributed homogeneously, but rather concentrated in certain areas. These areas can be described as *frontiers* – places or regions with specific land-use dynamics leading to the rapid development of the exploitation of some land or resource, and that experience marked social-ecological transformation accompanying and resulting from resource exploitation.

*Land-use frontiers* include agricultural frontiers, where the resource is land suitable for agriculture, but also fit a broader set of land uses (mining, forestry activities, energy production, conservation… as well as associated resources such as water). Frontiers can expand into natural ecosystems or already converted land. Frontiers can occur through so-called crop booms, which are sudden, rapid, and intense transformations from one dominant type of land-use to another, generally a commercial crop (3,4). Frontiers appear in multiple contexts and exhibit a variety of spatiotemporal patterns. As political, technological, economic and environmental – e.g., climatic – conditions change, new frontiers emerge, such as new agricultural frontiers in northern (boreal and arctic) regions (5,6). Beyond their environmental impacts, frontiers are also hotspots of profound socio-economic change, impacting livelihoods, food security, development, social structures, cultures, and Indigenous People and local communities' rights (7,8).

*Deforestation frontiers* characterize places with rampant conversion of forest ecosystems, generally related to agriculture (9), but agricultural frontiers can also expand into other, non-forest ecosystems such as grasslands and wetlands (10). Current deforestation frontiers are largely located in the tropics. Over 50% of tropical deforestation in recent times has been concentrated in twenty-four deforestation fronts (11). Across tropical dry forests and woodlands, about a third of deforestation occurs within landscapes characterized as frontiers (12). Global scale models and assessments using aggregate agricultural area or production data can easily misrepresent dynamics in deforestation frontiers, as these areas represent a small percentage of global area and production. Indeed, over a global agricultural land extent of ~4745 Mha (~1562 Mha of cropland and ~3183 Mha of pastures) (13), the annual rate of agricultural-driven tropical deforestation is around ~6.4–8.8 Mha/y (9), thus corresponding to only ~0.1-0.2% of the global agricultural land base. Assessments and models of food production, yields, and even land use area can thus be highly accurate overall while grossly misrepresenting the specific dynamics of deforestation frontiers.

Further, adequately understanding, explaining and possibly modeling the dynamics in land-use frontiers requires contextual, focused studies that acknowledge the complexities and multiple factors that shape them. Simple, universal logics such as based on economic theories alone might be insufficient. Indeed, at least a third of agriculture-driven tropical deforestation happens without land being subsequently used for recorded agricultural production, due to messy dynamics involving speculation, tenure ambiguity and disputes, conflicts, mismanagement or institutional failures (9,14). Global statistical models based on standard land-rent theories (relying mainly on accessibility and agro-environmental qualities of the land) can explain the broad patterns of global cropland extent and distribution, but reveal vast heterogeneity in how these factors play out (15), and struggle to explain recent global cropland expansion in frontiers (16). Further, past frontier dynamics partly explain recent dynamics, suggesting some forms of path dependence (16).

As we will show, much has been written to describe and explain the functioning of active frontiers, especially those related to deforestation and agricultural expansion. Yet much less research unravels how these frontiers initially emerged in territories considered marginal in terms of productivity and/or connections to global markets (17). Our objectives are (i) to synthesize the conceptual and theoretical tools that can describe (Section 2) and explain (Section 3) frontier processes, and (ii) to build on these to propose an integrated theoretical understanding of frontier emergence (Section 4). We then discuss implications of our proposed theory for the governance of frontiers for enhancing sustainability (Section 5).

# 2. Describing different types of frontiers

Broadly speaking, frontiers correspond to spaces, places or regions characterized by rapid development of the exploitation of some resource and experiencing marked transformation accompanying and resulting from this exploitation. Academically, the notion of frontiers on which we build emerges in the works of the American historian Frederick Jackson Turner. In his work *“The significance of the frontier in American history”* (18), the “frontier” refers to the waves through which colonists and so-called civilization progressed across North America in the 19th century. In Turner's view, these material changes were intrinsically linked with and mutually influencing other societal processes including economic development as well as cultural, institutional and political changes. Turner’s main thesis is now largely criticized, but also rediscussed and reinvented (Section 3.1). Yet, a core intuition remains, namely that the notion of “frontier” can help understand the linkages between expansion dynamics and broader social and environmental change.

Land-use frontiers can be characterized in many ways, based on the actors that drive the process, the type of land-use dynamics and the context in which they unfold, distinct stages (emerging, active, post-frontiers), the types of expanding land-use (agricultural frontiers, frontiers of forestry or conservation activities), and others. As this is often a point of confusion when discussing frontiers, we unpack here some of the terminologies and typologies that have been developed to characterize land-use frontiers, focusing on four dimensions: the dominant land cover at the time of frontier emergence; the actors involved in the frontier; the land uses driving the frontier; and the stages of frontier development.

## 2.1. Land cover: primary versus secondary frontiers

Several prominent conceptualizations of land-use transitions rely on successive “stages” of land use (19–21). Initial or *primary frontiers*, or *primary conversion*, are places where land use - mainly agriculture - expands into land that is covered predominantly by natural vegetation that has not been converted to another use in the recent past, although it may be under some form of management or use by humans. Primary frontiers are often referred to as *deforestation fronts* or *pioneer fronts*. These primary transitions give rise to *secondary lands*, referring to lands that have been already converted at least once in recent times (22). In turn, *secondary frontiers* or *secondary conversion* refers to places where a new land use expands into secondary land, such as cropping agriculture expanding into cattle ranching or pastoralist rangelands, or a commercial crop expanding into areas used by semi-subsistence smallholders. This distinction is not straightforward or absolute, as most areas of natural forests or other natural ecosystems have been inhabited and used by indigenous people for a long time, and are thus characterized by gradients of intensity of use (23–25). Yet, in practical terms, the human and environmental implications of frontiers expanding by conversion of natural or semi-natural ecosystems versus those expanding into already converted land can be vastly different (26).

## 2.2. Actors involved: colonists versus corporate frontiers

Many studies have distinguished frontiers based on the types of actors that drive land-use change. These distinct types of actors and dynamics relate to vastly different underlying processes and resulting impacts. When frontiers are driven by the inflow of new actors into relatively low-population density areas, two broad categories emerge. First, *frontiers of settlement* (27) – also referred to as *colonist frontiers*, *populist frontiers* (28) or *smallholder frontiers* (29) – are driven by large numbers of settlers or colonists, who arrive from more populated areas in search of farmland. Settlers typically practice small-scale, semi-subsistence agriculture, including shifting cultivation or livestock husbandry, and their main asset is family labor (29). These settlement frontiers have been amply documented e.g. in the Brazilian Amazon (29–32), India, Bangladesh, the Philippines and Indonesia (27). Colonists can be internal migrants, moving from core regions of the country that are typically more densely populated, where land uses are more consolidated, and which are more tightly integrated within the political institutions, as in the case of migrants from Southern Brazil to the Amazonia, or from Indonesian Java to Kalimantan or Papua Islands, or Kinh migrants in Vietnamese highlands. But colonists can also be from abroad, such as European settlers in American, African or Asian colonies. Historically, governments have played a central role in populist or colonist frontiers by planning or fostering their development, for instance through the construction of infrastructures like roads, setting up migration programs such as the Transmigration programs in Indonesia (33), or supporting colonists financially and politically. As discussed in Section 3.6, the state's motivations for initiating or supporting such settlement frontiers can encompass a range of aspects including geopolitical (territorial consolidation and securing borderlands), economic (fostering economic development and foreign earnings through commodity exports), or internal political stability motivations (mitigating potential social unrest from unequally-distributed economic growth or population pressure in core regions of the country).

A second broad type of frontiers is driven by investors (corporations, companies, commercial farmers and other capitalized and market-oriented actors) operating medium- to large-scale commercial farms, cattle ranches or industrial plantations. This second type of frontier has been called *corporate frontier* or *corporatist frontier* (28). Similarly, Pacheco (29) distinguishes between the Bolivian agricultural frontiers dominated by smallholder farmers and landscapes dominated by large-holder cattle ranchers in the greater Amazon that produce a marketable commodity (beef), labeled as *capitalist frontiers*. The term *commodity frontiers* was introduced by Moore (34) to refer to frontier processes that operate for and through the development of a land-based commodity (e.g., timber, sugar, tobacco, cattle, and other frontiers), integrating previously largely disconnected regions into the world economy, and often requiring large amounts of capital. These frontiers experience population inflows but typically much less than populist frontiers, with capital being the primary production factor flowing in. Recently, governments have shifted from a role of planning to one of facilitation (35) or of nonintervention, giving rise to the notion of *neoliberal frontiers* (36,37), where export-oriented corporate farming is more strongly motivated by global demand and deregulated access to land than by government subsidies. From that, the term commodity frontiers was also used to describe areas where the expansion of the production of agricultural commodities (e.g., beef, soy, or palm oil) by large-scale farms is shaped by the greater ability of these actors to influence and capture economic rents (38).

## 2.3. Land uses: Logging, mining, conservation and restoration frontiers

Much of the literature and reasoning exposed in the previous and following sections have been initially formulated with a focus on deforestation and agricultural frontiers. Yet similar frontier trajectories have been described for other land uses, such as timber (39,40), including waves of successive logging of trees of different qualities in natural forests (41), or for frontiers driven by extractivist or energy activities such as mining (42), oil and gas exploitation (43), hydropower construction (44), or the expansion of solar and wind power (45).

Research has also described *eco-frontiers* (46), *green frontiers* (47), and *conservation frontiers*, where conservation actors target land for conversion into a protected area for its perceived biodiversity, wilderness, or ecological importance (48–51). This can happen in remote, marginal lands, but also in lands used or targeted by other actors. Consequently, conservation actors may actively compete with other actors and land uses, be they long-standing, such as local communities, or new frontier agents (e.g., incoming migrants or agribusiness investors). Similarly, *forest restoration frontiers*, where restoration goals or carbon forestry have replaced commodity expansion as the main driver of the frontier, are playing an increasingly important role in land system dynamics (de Jong et al., 2021). All these frontiers can be apprehended through the frontier theories discussed in this article.

## 2.4. Stages of frontier development

Research on land-use transition has suggested “stages” of frontiers (pre-frontier, post-frontier…). Frontiers typically develop in a non-linear way, with take-off or *frontier emergence* that can appear as "abrupt" or "surprising", then accelerating towards an *active frontier* with rampant land-use dynamics (38). Although these can be useful to characterize distinct dynamics (52–54), one should not assume that frontiers necessarily move predictably through these stages (for a discussion, see (55)). Rather, these stages mainly help to describe distinct combinations of actors, drivers, land-use dynamics, and impacts. With this in mind, intensive research efforts have been focused on describing and explaining the functioning of active frontiers in relation to agricultural expansion (12,14,56). Less focus has been put on explaining the processes that condition and shape *emerging* frontiers in territories considered as marginal in terms of agricultural productivity and global market connections, but that are in the process of turning into rapidly expanding active frontiers (17).

# 3. Theories on land-use frontiers

Multiple theoretical perspectives shed light on distinct aspects of frontiers (57). In a historical and etymological analysis of the word "frontier" (in French), Febvre (58) already identified many of the tensions of the concept, i.e. between frontier as a hard boundary that roots a political space into a territory, versus something dynamic and expanding e.g. under military force; frontier as a separation between different populations or, at the contrary, the place where these people meets; or frontiers as contested places versus marginal, neglected places.

In this section we unpack the main theories developed to explain frontiers. From the original notion of frontier as a tidal wave of colonists and civilization (18), frontiers have been described as a process of pushing back "wilderness" to create a space for development by taming the natural world (59), as well as spaces facing a rapidly expanding force that brings opportunities for a number of people (60). Frontiers have also been framed as places of resource extraction or exploitation (61). The "resource" in these frontiers can be either newly discovered or "reinvented", for example if it acquires a new value due to technological, institutional, socio-economic, environmental or cultural changes (38). This makes frontiers typical spaces of territorialization, i.e. spaces where institutional actors, including governments and corporations, turn places into "territories" that they can understand, monitor, regulate, and exploit (62). Through these processes, frontiers are also places of interface and friction between different worlds, e.g. subsistence and capitalist economies, different cultures, socio-political systems, and mode of relations to nature (63,64).

## 3.1. The tidal frontier

For Turner (18), in a context of settler colonialism (65), the expansion of the frontier and the rolling back of wilderness was an attempt to make livable space out of an uncooperative nature. This process, which is seen as unfolding as a tidal wave, was more than simply a process of spatial expansion and the progressive taming of the physical world. For Turner, the development of the frontier was thus not only critical for the development of the country in economic and political terms, but also the central experience which defined the uniqueness of the American national identity and values. Each new wave of expansion westward, in its conquest of nature, sent shock waves back east in the democratization of human nature (66). Turner's thesis has now been largely criticized, including for its erroneous and harmful vision of land being "empty" or "unused" (see (67)), justifying colonization and eviction of indigenous people, and its teleological association between the frontier process and the supposed unique character and value of the U.S.A. and its people. Yet, Turner's frontier conceptualization nevertheless included seeds for many of the subsequent theoretical developments presented below, such as the notion of successive waves of frontiers (the pioneer, the settler, the urban, see Section 3.2), how land uses in frontiers are influenced by local contexts (the ranching frontier in the Great Plains versus the mining frontier in the mountains), and how the frontier is not a thin boundary but rather a dynamic space, creating opportunities for some, in which different worlds encounter with frictions (Sections 3.8 and 3.9), and where states and other powerful actors deploy efforts to make the territory legible and assert control of it (territorialization, Section 3.7).

## 3.2. Frontiers as successive waves, and interstitial frontiers

Many theorizations emphasize that frontiers do not manifest as a singular process happening once and for all in a certain space or region, but instead as successive *waves*, which may build on each other, reverse each other, and often overlap. The relations between these waves can be seen as contingent or following a regular, predictable succession pattern. For example, the making of a “second” new resource and commodity frontier in the 20th century in Laos builds on a "first frontier" in French colonial time which profoundly transformed landscapes, property relations, institutions, and the development trajectory, paving the way for this second recent frontier (68). Hirsch (69) describes, in Thailand, the succession of an agricultural frontier, which came to an end after the 1990s, by a peri-urban frontier, both successively transforming landscapes and institutions.

The broad theory of land-use transitions corresponds to a typical conceptualization of frontiers occurring in predictable, regular waves. This theory sees land use following regular sequences from natural ecosystems to extensive, smallholder and subsistence land uses, then to intensive agriculture and forestry, then urbanization and the progressive rise of protected areas (20,21) (see section 2.2). Some deterministic versions of forest transition theories – i.e., large-scale shifts from deforestation to reforestation – articulate a similar sequence (55). Other theories such as the capitalist penetration theory (Section 3.2) also posit regular sequences from smallholder frontiers to consolidated, large-scale hollow frontiers.

Further, in von Thünen’s land rent theory (see section 3.4), concentric circles of distinct land uses progressively expand or contract depending on changes in equilibriums of the costs and benefits of these different land uses. Land rent theory has been widely used to explain successive frontiers waves expanding in a region such as in timber exploitation (41) or across sequences of land uses, such as the sequence from logging to cattle ranching to intensive soy cultivation frequent in South America (70,71).

The apparent "closure" of a frontier, i.e. the cessation of the expansion of the land use that was driving the frontier, does not imply that there is no more resource to exploit in this area, or that extraction activities will not resume in the future. Contextual changes can create the potential for renewed rent extraction by new actors arriving with previously absent capacities to create and extract rents (e.g. with new techniques or more capital than previous actors), or as new land uses trigger a new frontier (72) (see section 3.8). Similarly, the investigation of successive waves of investments in a frontier region in Northern Mozambique showed that, even when failing, early waves left *legacies* that might progressively build the conditions for another frontier wave to emerge, for example in terms of business approaches and ways for investors to deal with land conflicts, institutions and policies, land with legible tenure (see section 3.6), brownfields (land already cleared and with infrastructures for agriculture), financial capital, and social networks (17).

Several authors also contend that frontiers, rather than unfolding over the whole landscape like a blanket tidal wave as described by Turner, typically progress by hybridation, i.e., creating spaces where the former and incoming land uses, as well as actors and institutions, overlap (73). These hybrid landscapes create the opportunities for *interstitial frontiers*, where early frontier waves leave behind interstitial spaces–such as forest remnants, uncultivated land, or pockets of extensive or subsistence land uses that are neither unmodified nor completely transformed. In further waves, these interstitial spaces are progressively contested, as seen in the multiple forms of extractive activities in Indonesia (73), or in Palestine's West Bank, where interstitial spaces of land that was previously either uncultivated or used non-intensively suddenly become the object of intensive agriculture and struggles for water control (74). These authors argue that much of the interactions and frictions (Section 3.7) that happen in frontiers as places of encounter between different worlds happen in these interstitial spaces rather than at the supposed edge of a frontier wave (75).

## 3.3. Theories of colonist frontiers

Several theories have been formulated specifically to describe the processes of settlement, colonist or smallholder frontiers. A classic theory linking populist and corporate frontiers is the *capitalist penetration thesis* (31), which sees frontiers as occurring through waves where, after the initial populist, smallholder settlement of the frontier, the land is then progressively consolidated by capitalized, corporate actors who switch land uses from mixed, semi-subsistence farming to simplified, large-scale commodity production, such as cattle ranching or soybean. This gives rise to depopulated *hollow frontiers*, while the initial colonists are pushed further away to the periphery, perpetuating the frontier process in more remote regions (36,76,77). A complementary, *inter-sectoral articulation thesis* argues that smallholders at the frontier serve a functional role in urban economic development by providing low cost food (subsidized by unpaid family labor) and serving as a reserve of cheap labor that can be activated by inflows of capital whenever there is a surplus of capital in the core, urban economy (31). The term *dualistic frontier* acknowledges that in a number of real-world contexts, both populist / smallholder and corporate / commodity actors operate jointly to produce frontiers (19).

A distinct set of theories, building on *Chayanov's peasant theory*, focus on how the *life-cycle of smallholder households* produces the frontier process (31,78,79). These theories argue that smallholders initially settling in the frontier, typically young parents with non-working, dependent children, will focus on securing harvests on a relatively small land base, with minimal risks. This triggers an initial expansion spur, which has to be periodically renewed as soil fertility of cultivated plots declines. As children enter the workforce, household vulnerability decreases and they become more risk taking, typically expanding the land base and engaging in more risky but potentially more rewarding cash or commodity crops. The children subsequently establish their own household, furthering the frontier process. Older households with lower labor force would then turn to less labor-intensive activities such as cattle ranching. These successive cycles drive much of the spatio-temporal patterns of frontier expansion. Yet, various demographic and socio-economic household and contextual factors lead to more complex relations between household life cycles and frontier dynamics (78–81).

## 3.4. Resource frontiers and land rents

A central conceptualization is that of *resource extraction frontiers*, frontiers as places of imbalance between abundant natural resources and a comparative lack of production factors (i.e., capital, labor) to exploit these resources (19,38,57). Frontiers then form due to the rapid inflow of production factors (labor, capital) and the associated increase in resource use rates as actors seek to capture economic rents. Rent theory has emerged as a dominant explanatory framework in economics for understanding frontier expansion and development (30,32,82,83). It posits that rapid agricultural expansion is driven by the economic rents that actors capture in situations of land abundance. Rents influence land-use decisions such as whether and when to move into the frontier (84), where to invest, what to produce (32), and how much land to clear (30).

The classic *land rent theories* are those of von Thünen and Ricardo (57,85–88). In these, the land rent is determined by accessibility and biophysical land quality, respectively. Von Thünen's theory proposes that land uses typically organize in concentric circles around a central place. Land rent theories build on the distinction between the actual economic rent that can be extracted from land, and the bid rent, i.e., the value that economic actors are willing to offer to acquire or use the land. Explanations on bid rent have relied on both an "equilibrium" and a "disequilibrium" view of land rent (89). In the equilibrium view, the bid rent is aligned with the economic rent. With changes in accessibility, quality (or technologies allowing to use certain types of land more productively), or market demand, the area on which certain land uses can generate a positive rent expands or contracts. Thus the frontier progressively expands along with improvements in roads, techniques for improving soils or adjusting crops to certain agro-environmental conditions, or increases in demand.

The equilibrium perspective can explain the longer-term development of frontier areas, but understanding the sometimes very abrupt, possibly surprising, dynamics in many frontiers requires to adopt a "disequilibrium" perspective, based on the notion that production factors cannot instantaneously adjust to equilibrium changes, and that not all actors are equally positioned to induce changes in economic rent or capture these rents. Along this line of thought, 'early frontiers' emerge when actors identify and capture 'abnormal', high rents created by a gap between the economic rent and the bid rent, so that previously uninteresting areas become potentially valuable (38). Five main mechanisms may create sudden rise in abnormal rents: 1) improved accessibility; 2) changing agro-environmental conditions (e.g. increase in rainfall); 3) new technologies (e.g. genetically modified soy); 4) a rise in producer prices or demand; and 5) favorable policies.

The sudden rise of rents is explained differently for populist and corporate frontiers. In populist frontiers, rents are primarily created by (authoritarian) states (32). In Brazil, in the 1970s and 1980s, the migration and settlement of poor and landless farmers in the Amazon was planned and organized by the state through colonization schemes (31,90). Even when migration followed a seemingly more spontaneous trajectory (30), state interventions such as road construction into hitherto inaccessible areas and agricultural or settlement subsidies (32) made it possible for struggling farmers to move (38). In capitalist, corporate frontiers, rents are consolidated through markets, while the state plays a more indirect, though not unimportant, role through a different set of neoliberal policies (32). The cattle frontier in the Brazilian Amazon for example, was driven by the rising global demand for beef in the early 2000s, but corporate ranchers could not have captured these market-generated rents were it not for the state to have continued building roads (reducing transportation costs), launching a campaign to eradicate foot-and-mouth disease (improving animal health and product quality), and introducing monetary and trade reforms that favored export production and market liberalization during the previous decade (82,91). Yet, beyond the role of the state, large-scale actors that drive corporate, commodity frontiers can have strong agency to shape and capture potential abnormal rents compared to smallholders, who have less agency in terms of access to new technologies, influence on policies, or infrastructure development. These large-scale, capitalized actors are well-positioned to shape and capture these rents, and often outcompete small- and medium-scale farmers in contemporary corporate frontiers (38). Yet, depending on a number of factors such as resources' ecological characteristics, inputs, labor, and knowledge requirement, or the technical characteristics of downstream processing, smallholders may also have a comparative advantage, as smallholder-driven Southeast Asian frontier booms exemplifies (92).

## 3.5. Agglomeration economies in frontiers

*Agglomeration economies* correspond to economies of scale external to an individual company, but internal to the sector (93,94). Agglomerations create localized clusters of specialized knowledge, inputs, and industry-specific infrastructure and institutions. The process of clustering lowers transaction and production costs, promotes learning and innovation, increases local competition, and enables leveraging collective political agency (93–96). A central feature of agglomeration economies is that they are dynamic, and they build up with the density of actors of a certain sector in a certain area (93). Therefore, in early stages of frontiers where the expanding land use only occupies relatively small areas in the region, actors are confronted with low levels of agglomeration (97). With the expansion of the frontier, agglomeration economies are expected to build and to support increased profitability and productivity of farming and other activities (98,99).

Theoretically, one therefore expects an inherent, but not perfect, degree of trade-off between the possibility of extracting rent from resource frontiers and the presence of agglomeration economies (97). Resource frontiers and agglomeration economies encapsulate the compromise an investor makes in choosing cheap land with the potential to expand and achieve scale economies internal to the firm, as opposed to locating closer to an investment cluster to benefit from the existing scale economies external to the firm. Actors may also seek a compromise, i.e., places that still have sufficient resources for extracting rent while having already a certain amount of agglomeration economies to sustain operations.

## 3.6. Frontiers as places of accumulation by dispossession

The process of rent creation and capture described above (Section 3.4) has been theorized with the Marxist notion of *primitive accumulation*, i.e., the processes through which "initial" appropriation of capital, in the form of natural resources or labor force (slavery), occurs (100). This notion has evolved into the concept of *accumulation by dispossession*, i.e. the concentration of capital and resources in the hands of a few by dispossessing the public sector and the bulk of private actors, including through the establishment of property rights and the support of neoliberal policies (100,101). Multiple forms of claims-making, through exclusion, expropriation, and enclosure, violent or not, underlie such accumulation of land by dispossession (7,102,103), and contribute to the dynamics of capitalist penetration (Section 3.3), neoliberal and corporate frontiers, and creation and capture of abnormal land rent by corporate actors (38). Smaller-scale actors can also sometimes play as agents of dispossession and land-grabbing, involving micro-level processes that mirror those that characterize large-scale land grabs (92). Dynamics of accumulation by dispossession are, in principle, led by private actors, but they are nevertheless strongly interlinked with the role of states and subnational governments in frontiers. On the one hand, states can be actively engaged in or support this accumulation by dispossession, which can be intrinsically linked to dynamics of territorialization (Section 3.7). On the other hand, as frontiers are typically at "the edge of the state" (104), i.e., at the margins of states' reach, the mere neglect or lack of control of remote frontier regions by state authorities, and the neglect by public opinion in distant consumer markets and environmental organizations, can suffice to leave the ground open for powerful actors to concentrate land and resources (105).

## 3.7. Frontiers as spaces of territorialization

Another perspective put forward to explain frontiers is the idea of frontiers as spaces of *territorialization* (62,106). Territorialization is the process through which states or other actors, e.g., large-scale corporate actors, turn a "space" into a "territory" – something that is legible, can be surveyed, monitored, recorded, administered, taxed, and governed. In this perspective, frontier spaces are "*transitional reconfigurations of institutional arrangements*" (62). Frontiers are thus sites of contentious encounters over authority and redefinition of institutions and social contracts.

Territorialization, growing commodification and corporatization, and accumulation by dispossession (see Sections 3.2, 3.3) can proceed together, as in the case of the Ethiopian pastoral frontier (107). Korf et al. ((107)) rejoin Peluso and Lund's (102) observation that there is not one grand land grab, but a series of changing contexts, so that the Ethiopian pastoral frontier is not primarily driven by outside forces but by 'indigenous' capital and entrepreneurship that include investors from the diaspora and political "big men" (Somalis inside and outside of Ethiopia). As part of the territorialization, the emergence of land markets, either based on formal or customary land laws, renders the land tenure legible for private actors–a crucial aspect that contributes to the expansion of commodity frontiers and rent appropriation (108).

Nation-states are often interested in enabling such territorialization processes for larger political purposes such as integrating marginal or peripheric territories, consolidating borders and establishing geopolitical claims over spaces and resources, as well as reducing population pressure and social unrest in "core" regions of the country. This has been abundantly shown in contexts such as in the Brazilian Amazon (30), the Chaco region in South America, or in Southeast Asia – Vietnam, Thailand, and transmigrant programs in Indonesia (106,109,110).

## 3.8. Frontiers as places of frictions

Frontiers are also seen as places of *friction*, liminal places that are at the interface between different worlds, cultures, modes of relating to nature, patterns of environmental exploitation, or socio-political systems, e.g. subsistence and capitalist economies (63,64,111). This coming together of different worlds can cause frictions to arise. These frictions can manifest as ethnic and sociopolitical conflicts at the margins (104,112), such as forms of environmental exploitation (111) or spatial conflicts between an underprivileged minority and the settler group which represents and is represented by the hegemonic national power (113). Thus, whereas land conflicts in commodity frontiers are often conceptualized as the local consequences of global political-economic changes (34,114,115), a frictions perspective offers a more nuanced view – namely that land conflicts rarely correspond to mere projections of global dynamics, but are instead best understood as resulting from complex interactions involving local and global dynamics (116).

Conflict and friction are also often linked to processes of abnormal rent capture and dispossession by accumulation, as actors struggle for claims and control of resources (Sections 3.4, 3.6). Frictions can also arise when the state and other actors engage in territorialization to change the readability, governability, or exploitability of frontier spaces (Section 3.7). Yet, as mentioned above, frictions extend beyond the political and economical realm and arise also from the encounter between different languages, communication modes, production systems, or worldviews. They often reflect tensions to redefine the vision of what the frontier place is meant to be. For example, the act of imagining the territory of Mozambique as a blank slate to be exploited and transformed is an act of historical erasure of alternative visions of its history (117,118). For Turner, the westward spatial expansion of the US frontier was associated with the forging of the US's national identity (66). These frictions produce conflicts but also lead to reconfigurations and the emergence of new institutions and cultures.

## 3.9. Frontiers as spaces of opportunity, anticipation, and imitation

Frontiers also connote a *space of opportunity* (60). In that view, the "resource" from resource frontiers theory can be seen more broadly as any untapped potential to fulfill certain objectives, for which the actors transform the place. In that view, *"a 'frontier' refers to a place that is facing an expansive force"* (60) in a rapid and overwhelming manner. This can be the penetration of a new technology or product into untapped markets, but also the expansion of a religion, a political ideology, a technology such as electrification, etc. For example, African cities have become particularly attractive to the global property development sector, and are often referred to as the world's 'last frontier' for real estate developers. This view also necessarily implies that the frontier is transitory in space and time: At some point, the untapped opportunity is exhausted and the frontier, or window of opportunity, closes. Yet, it does not prevent another opportunity from arising and restarting a new frontier (Section 3.2).

Frontiers as spaces of opportunities can also be analyzed through the notions of *expectations* and *anticipation*. Anticipation is about expecting something to happen and, importantly, doing something about that expectation (119). In Northern Mozambique's land-use frontiers, many of the land-use decisions and practices have been anticipatory and shaped by expectations of future profits, speculations about resource potentialities and dreams of better futures (120). Similarly, in northern Laos' agricultural frontiers, the expectation that a stable rubber market would develop drove farmers to adopt the crop long before such a local market materialized (4). Anticipation thus shapes land-use change, particularly in emerging frontiers where the future is uncertain yet hopeful. Anticipation and expectation can fuel and be fueled by what Anna Tsing terms the *economy of appearances* (121)–the conscious making up of economic success to attract investors to a business idea even when, factually, sustainable economic gain is missing. Frontiers as spaces of opportunity lend themselves to such performance, as exemplified by the spectacular rise and fall of the mining company Bre-X claiming to have tapped into very large gold deposits in Indonesia, which never materialized (121). As a counterpoint to expectation fueled by appearances is the spread of information based on tangible, economic success, for example of early innovation adopters. Such early successes can trigger processes of *imitation* that become self-reinforcing and can lead to boom-like expansion of agricultural crops (122).

# 4. Proposing an integrated theory of land-use and resource frontiers emergence

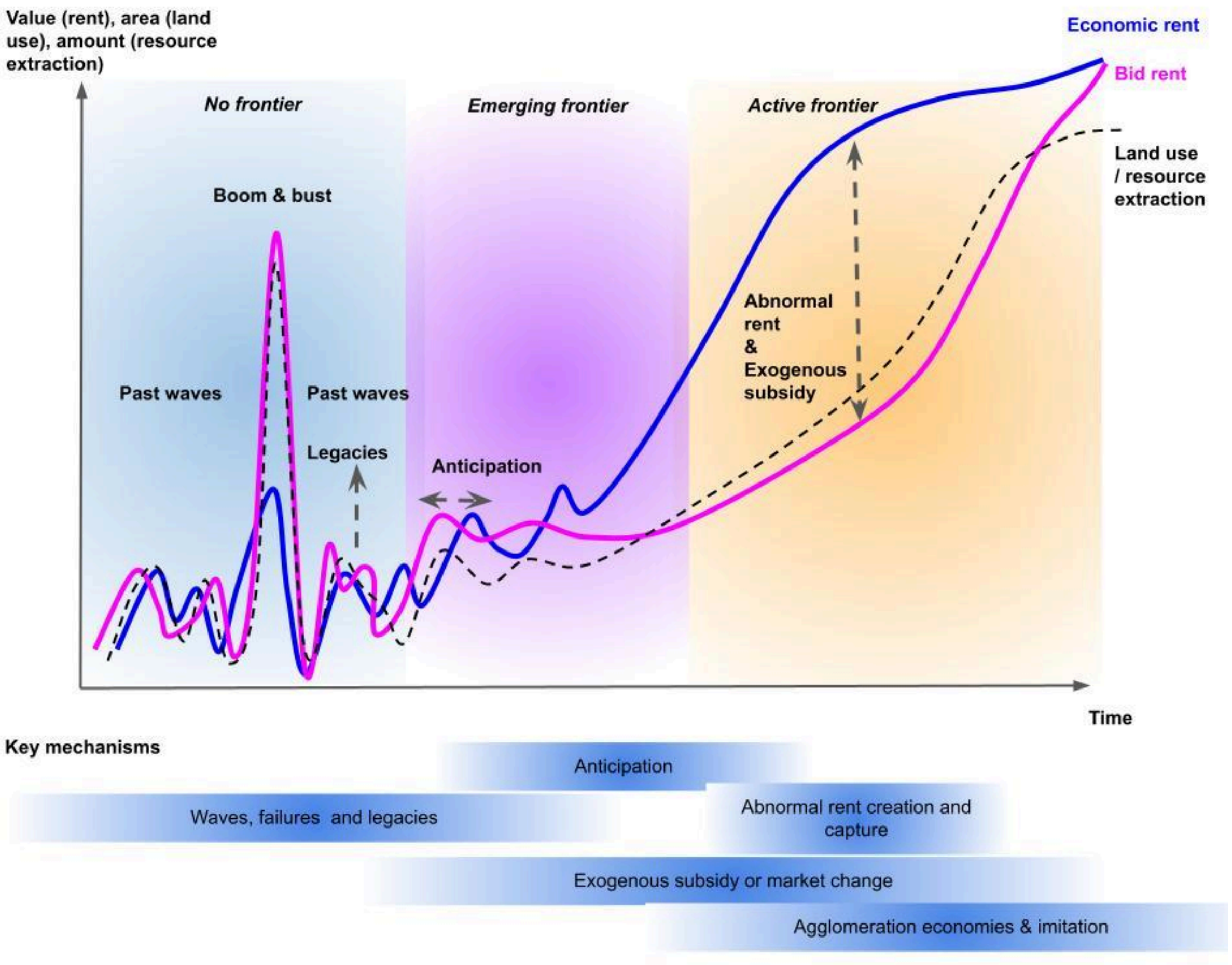

***Figure 1. Our proposed frontier emergence theory.*** *The starting point is an area that has the potential to become a resource frontier, but is not yet one. The extent of land use and*

*resource extraction fluctuates over time, as well as the economic rent (the value that can be extracted from this land) and the bid rent (the value that actors are willing to pay for the land). Under several key mechanisms, the area might become an emerging frontier and then an active one. Even though most mechanisms operate at the different stages (i.e., no frontier stage, emerging frontier stage, and active frontier stage), the theory proposes that certain mechanisms are key at certain stages. The theory is not deterministic, i.e., it does not say that the area will necessarily go through these stages. The "no frontier" stage, which can be characterized by multiple waves that do not trigger a large-scale frontier, can last for a long period, and other trajectories are possible.*

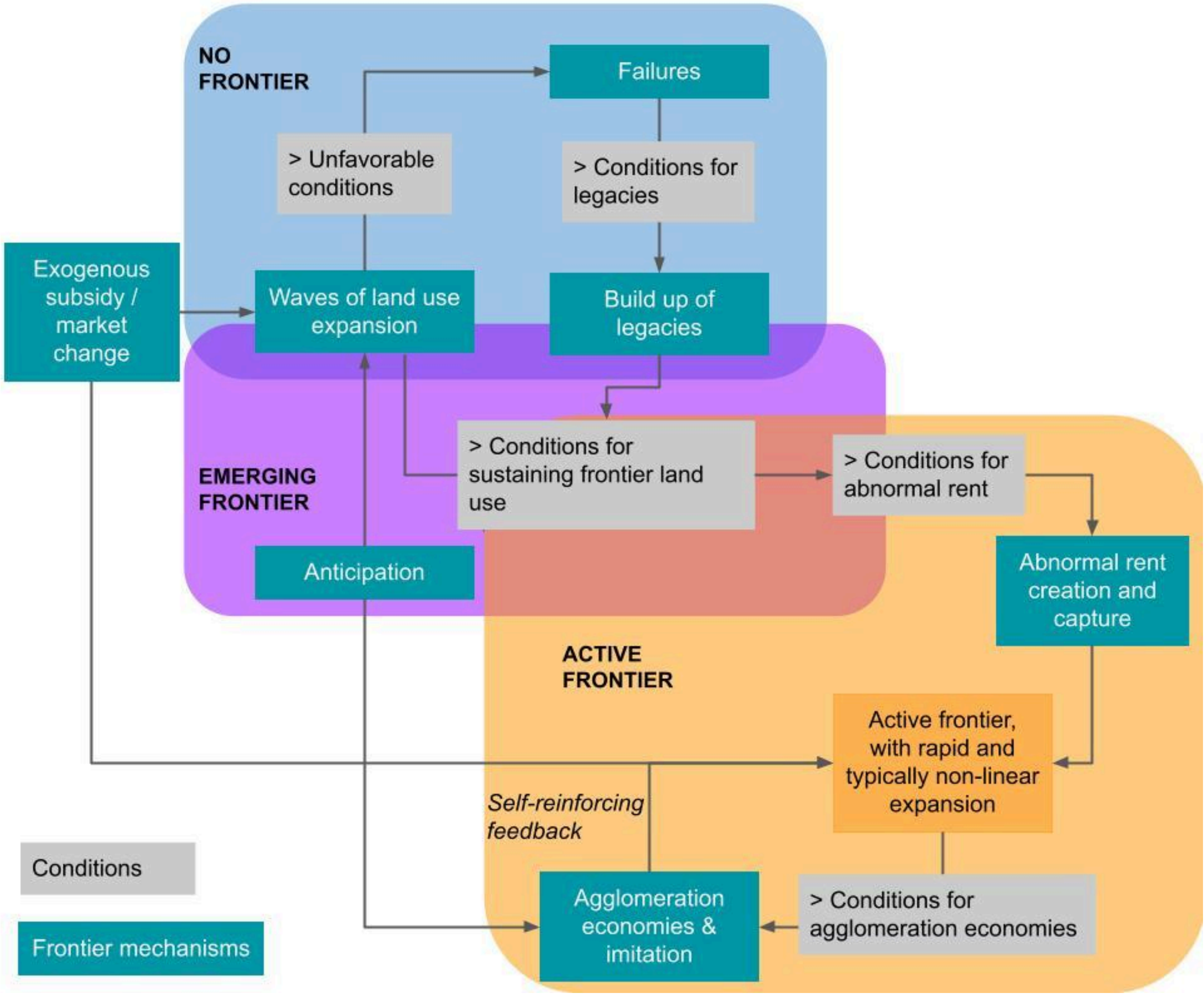

***Figure 2. Mechanisms and conditions in our proposed theory of frontier emergence and activity.*** *The starting point is an area that has the potential to become a resource frontier, but is not one. Through several mechanisms that operate under certain conditions, the area might become an emerging frontier and then an active one. Economically, "unfavorable conditions" correspond to negative economic rent, "conditions for sustaining frontier land-use" to positive economic rents, and "conditions for abnormal rent" to economic rents being largely above bid rent at least for some actors. But the conditions encompass more than economic factors. Only key mechanisms are represented here. The theory is not*

*deterministic, i.e., it does not say that the area will necessarily go through these stages, and other trajectories are possible.*

Several theories exist about how already active frontiers unfold yet fewer works have proposed explanations of how frontiers do emerge or not. Here, we aim to put together these distinct pieces to propose a theoretical account of what happens before frontiers emerge, when they emerge, and when they are active. We build on the notion of middle-range theory, understood as a *"contextual generalization that describes chains of causal mechanisms explaining a well-bounded range of phenomena, as well as the conditions that trigger, enable, or prevent these causal chains"* (57).

We structure this account by describing frontier dynamics as a set of stages (Figure 1), where distinct mechanisms and factors dominate (Figure 1, Figure 2). Not all areas are expected to experience frontier processes with a regular succession of these stages. Yet, there is a continuity in the chain of causal mechanisms across these stages, which are likely to occur over a certain range of conditions. We discuss some of the conditions below, but for some mechanisms, e.g. legacies, the characterization of these conditions remain incomplete and to be further researched. A focus on conditions helps explain why some regions with the potential for frontier emergence do not undergo such a process, if certain conditions are not fulfilled.

The first stage is when a region has the potential for frontier development, but is not in a frontier stage, i.e., it has an abundance of certain untapped resources and potential opportunities, but it is not experiencing rapid inflow of production factors and resource extraction. A coarse analysis, for example focusing on land-use conversion as seen from satellites or through production statistics, might show these regions as largely stable or static. However, a closer inspection might reveal multiple failed attempts by pioneers and other actors to put the resource into use, facing unfavorable conditions and obstacles to investment and production.

In some cases, these early investments form waves that can take the form of commodity boom and bust cycles, where the bid rent increases quickly as actors rush to acquire land with the prospects of high gains from extracting a certain resource or growing a certain crop. If the actual economic rent doesn't increase according to expectations, because the resource is not as abundant or valuable as previously thought (e.g., land is abundant but much of it is poorly suitable for a certain crop), because production costs are higher than expected, or because demand for the products does not materialize, the rush ends up being only speculative, and comes to a bust when resource extraction, economic and bid rent all go down again. This mismatch of anticipated economic rents with actual ones is frequent in commodity frontiers where information on resource quantity and quality as well as costs and output markets is uncertain and intransparent. These regions are prone to economies of appearances, where various actors try to trigger and raise others' expectations.

While these successive waves of investment and land rush may appear as some form of noise in the big picture, they are nevertheless key as they may progressively build a set of legacies in the form of business approaches, institutions and policies, land tenure legibility, brownfields, financial capital, social networks, and lessons learned from land conflicts with local farmers (17). In terms of rent, both the potential economic rent that actors might extract as well as the bid rent that actors may offer for land and other resources remains rather low, fluctuating with the successive waves depending on various factors (availability of capital, demand for products, actors expectations, etc). Yet, progressively, “under the radar”, the legacies of previous waves might accumulate and lead to a gradual, hard to notice, increase in the potential economic rent that some actors might gain.

At some point, the region might see the emergence of a frontier. How does it happen? This original emergence can occur through at least two different but non-exclusive mechanisms. On one hand, it can be driven by exogenous factors: there can be a strong external subsidy to start with (e.g. direct financial subsidy or through infrastructures building, opening roads through the forest…), typically by states, which support the cost of frontier establishment when, in the absence of agglomeration economies, land use has very low profitability and competitiveness. This happened in many settlement frontiers as discussed above. The notion of frontiers as places of territorialization is central here to explain how and why large external actors such as states or even, in some cases, corporations, would deploy such large efforts to overcome the constraints on frontiers kickoff. When led by private actors, these exogenous inflows of resources would typically either occur in periods of high capital availability and lower returns in other sectors (such as the global financial crisis in 2007-2008), or require what can be called “deep pockets” actors, i.e. corporations or non-profit organizations that can tap into a large pool of resource during a sustained period of time, with long-term objectives guiding their decisions (97).

Another series of mechanisms driving frontier emergence are more endogenous, having to do with anticipation (i.e., expectations and actions thereupon) by actors, either local or external but focused on that region. Building among others on legacies of earlier waves, pioneers may develop sufficiently strong beliefs that farming or other resource use can become profitable in the future to continue to invest even in the face of adversity and low profitability, and even attract more people that believe in their anticipation narratives, creating “cohorts” of investors (83). These activities might eventually drive the structural changes that create the conditions for abnormal economic rents to arise, or the number of people and the amount of activity involved may surpass a certain economic throughput threshold to shift from "economies of anticipation" to "economies of agglomeration". In a dynamic akin to a self-fulfilling prophecy, economic rents may catch up with bid rents and ultimately surpass them.

When economic rents rise well above the bid rent, the frontier process is activated, accelerating with the gap between the high economic (or “abnormal”) rent that some actors can create and capture, and the bid rent of land and resources (38). Exogenous subsidies, at this stage, can also function as a quasi abnormal rent, i.e. allowing for rent capture for at least some of the actors, namely those who can claim the subsidies. Such abnormal rents can be supported or exacerbated by various endogenous or exogenous processes, such as legacies of previous waves or processes of accumulation by dispossession. Eventually a certain critical mass is reached at which agglomeration economies take over and frontier expansion becomes self-reinforcing. Moreover, such self-reinforcing processes as well as the initial dynamics of anticipation can be further reinforced by other positive feedback loops, which are typically social, such as imitation, which feed into one another (122).

At some point the bid rent catches up with the economic rent, eliminating the disequilibrium and, from that, driving frontier expansion in line with changes in land rents as explained by standard equilibrium land rent theories. Ultimately, the frontier may start to stagnate or saturate, particularly if the depletion of the resource makes it scarce or its extraction more costly. At this stage, the region can be described as “consolidated”, but it might at some point experience the emergence of a new frontier caused by renewed anticipation or a sudden new source of land rent disequilibrium.

The process of frontier emergence can fail due to external factors like wars or economic crises, as well as endogenous dynamics, for example if legacies of past waves are insufficient to raise the economic rent, or failures depress actors’ anticipations and prevent a subsequent wave. Commodity boom and bust cycles can contribute to the depletion of assets and resources, for instance by degrading the land and making it harder for a further frontier wave to emerge. Conversely, boom and bust cycles might pave the ground for further frontiers by destabilizing land uses and social, economic and institutional structures: leaving behind an unprofitable or unmarketable crop that necessitates replacement by a different commodity crop, or changing local people’s attitudes into a “boom mentality” focused on “quick money” (3).

States and other actors' exogenous subsidies and efforts may be too small to overcome the endogenous constraints on resource use. Territorialization efforts might face too strong a resistance from local populations. If no abnormal rent is created, the frontier can proceed only through incremental adjustments of the equilibrium land rent, which is presumably more gradual than the capacity for certain actors to create sudden, abnormally high rents. Throughout the process, the role of uncertainties and asymmetric information is crucial, including in creating gaps between economic and bid rent (overestimating or underestimating the value of a resource). Actors’ anticipations are necessarily based on uncertain knowledge about the future. Abnormal rents are strongly linked to different actors having different levels of information (e.g., on quality of land for agriculture, timber or mining resources in an area, etc). The same decision taken in different circumstances can lead an actor to benefit from large rents when the economic rent surpasses the bid rent, in abnormal rents conditions, or to be doomed to failure if the economic rent is too low and that actor does not have the deep pockets to sustain investments over a long time.

This is a stylized account; real world situations might deviate from this account or borrow only some of its elements. We highlight here the key mechanisms that operate at each state, but they can also operate at other times: anticipation and expectations of actors influence their decisions all along the way; agglomeration economies can occur, even if only lightly, in early frontier situations, etc. This theory largely fits with distinct types of actors and land uses or resources, though with nuances. Frontiers dominated by smallholder colonists would typically be more supported by exogenous subsidies from governments, while in corporatist frontiers with powerful actors, the mechanisms of abnormal rent creation and capture might be more important.

# 5. Frontiers, sustainable development, and governance

This article focuses on *how* frontiers emerge, more than *whether they should emerge* or not. Whether frontiers are desirable for local communities, external actors or stakeholders, or sustainability in general is another question, which likely has no singular answer. While we do not respond to this question here, our frontier theory can provide insights for sustainable development in regions where frontiers do happen. How frontiers emerge and unfold, possibly fail, or finally transform to other land use dynamics has profound implications for the trajectories of rural development and sustainability outcomes in these frontier regions and beyond. Much of the early literature on frontiers in tropical, low and middle income countries had a rather pessimistic outlook about the fate of smallholders in colonist frontiers. Narratives have often revolved around how poor smallholders were opening frontiers, either spontaneously or with state support, to escape land scarcity and rising land prices in consolidated regions, yet arguing that their fate in frontiers was often not better, as their land use would lead to land degradation, soil exhaustion, poverty and failure (123). Further waves of capitalist penetration through powerful actors would then consolidate the land and concentrate resources, leaving smallholders with the only option of furthering the frontier process ((124), see Section 3.3). Yet, the evidence shows that different frontier trajectories can lead to vastly different outcomes. Some frontiers give rise to consolidated land systems that may allow conserving natural habitat, that maintain denser populations, or even have a positive distributional impact on land or economic flows. Some frontier regions may become economic “power-horses” in the country, such as Mato Grosso in Brazil. A theorization of how frontiers emerge and unfold, as proposed above, leads to five insights on how to steer frontiers in order to contribute to sustainability and equity.

First, several theoretical lenses on frontiers presented above converge to typically conceive frontiers as places that are empty or unused. Indeed, the emergence of many frontiers builds on narratives about the abundance of unused land and untapped opportunities, and the resource frontiers theory itself is based on this notion. Multiple efforts have been made to assess the extent of lands that are considered as “unused”, “underutilized”, “marginal”, “empty” or “uncontested” (67). Yet, ample evidence, backed by theoretical developments, make it clear that such lands are very rare. Most land is used somehow, even if for the purposes of being set aside, and thus, land-use frontiers almost always unfold on land that is already claimed, occupied, used, managed, or transformed by people, or already provide benefits (23,25,67). Traditional smallholder colonist frontiers often expand into land occupied by Indigenous people, as do many corporate or commodity frontiers, which also expand into areas occupied by smallholders, semi-subsistence or traditional land uses, including pastoralists systems. Recently emerging renewable energy frontiers also typically expand on land already used, although it can also “stack” onto these lands, for example through agrivoltaics (125).

In contrast, other frontier theories such as the capitalist penetration thesis or frontiers as spaces of territorialization and frictions are precisely based on the principle that these spaces are *not* empty when the frontier unfolds. Our theory of emerging frontiers, which builds on the notion of consecutive waves and their legacies, also highlights that frontiers do not emerge in a vacuum. This is not in contradiction with the resource frontier premise that the frontier emergence builds on the availability of a resource (land) targeted for some purposes. In frontiers, the resource is there, but it is often used for other purposes, at other rates of extraction, and/or by populations that don’t have formal tenure rights, or have rights but have a power differential with other actors that aim to claim these resources. Recognizing that frontier processes almost always impinge on resource use rights and access might help to acknowledge and mitigate potential conflicts and losers, in contrast with a “terra nullius” narrative.

Second, resource frontiers incentivize unchecked resource extraction and thereby typically create unfavorable conditions for efficient use of resources (19,61,126). This implies that even though the frontier land use might be more intensive than the ones it replaces, land users are incentivized to expand to claim as much as possible of the resource, rather than dedicate their production factors to intensify. This is especially the case in primary frontiers where new land users enter into the area and land use expands into not-yet-converted ecosystems[1]. Conversely, theories of induced intensification (57,127–129) show that, under the appropriate capital, technologies, knowledge and market conditions, strong constraints on land-use expansion in frontier regions can lead to intensification of agriculture, e.g. in the Brazilian Amazon (53,130–132). This implies that (i) governing frontiers for halting conversion of natural ecosystems and improving sustainability outcomes does not necessarily imply a tradeoff with economic development, (ii) assuming that intensification will occur spontaneously in frontier contexts, and that intensification – whether spontaneous or supported by policies – will lead to a closure of the frontier goes against what we know about how frontiers work. Instead of assuming that intensification will spontaneously occur and reduce pressure on land (133), controlling land-use expansion should go ahead or at least hand in hand with interventions to support intensification.

Third, although *exogenous* subsidies, incentives, and triggers can play a key role in driving frontier processes, our proposition shows that *endogenous* dynamics including legacies of past waves, actors' anticipations[2], mechanisms for abnormal rent creation and capture, and the progressive built-up of agglomeration economies, also matter and may even be the main drivers of frontier dynamics. The focus on exogenous drivers likely influenced the somewhat pessimistic outlook often put forth in classic frontier theories–namely that of frontiers essentially emerging from external pressures on land and where the absence of "virtuous" endogenous dynamics inexorably leads to degradation and impoverishment. Such narratives were dominant for a long time with regards to cattle frontiers, assumed to have low potential for intensification and sustainable rural development due to a lack of both motivation and means, trapping cattle ranchers into extensive agriculture, poverty, land degradation, and perpetual expansion of the frontier. More recent works have shown the importance of endogenous dynamics, including the role of medium-scale institutions (institutions between individuals and the state), such as agricultural cooperatives, in mediating the demands from individual farmers towards the state, establishing better negotiating conditions with input and output suppliers, and thus leading to more socio-economically positive frontiers development and more productive agriculture (134). This line of thought is pursued in works on agglomeration economies in soy (98,99,135,136). Overall, these works explain how under certain circumstances an initial frontier can become more intensified, profitable, and provide livelihood means to relatively dense populations.

Fourth, frontiers progress through different stages or steps, which correspond to different processes, and thus in which different forms of interventions can be more or less effective

[1] In secondary frontiers that experience crop booms where one crop replaces another in already established agricultural landscapes, like banana boom replacing paddy in Northern Laos (122), the resource (land) is likely seen as less freely available, as paddy lands are already clearly delineated, with secure tenure. In that case, incentives might favor relatively intensive land uses, though likely less than in regions where the frontier land use is already well established and consolidated. In these contexts, many aspects of the frontier theory remain valuable though, including related to abnormal rents, spaces of opportunities, frictions, anticipations.
[2] Noting that not all actors with anticipations are endogenous, external actors can also form anticipations about the frontier region, and these two groups of actors can reinforce each other's expectations and anticipations.

(12). The proposed theory identifies emerging frontiers as a pivotal time, when the fate of land use and human-environment systems can still be steered through several levers (137,138). At this stage, mechanisms of abnormal rent capture and agglomeration economies are not strong and thus vested interests that benefit from a rolling out of the frontier are not yet powerful. As exogenous pushes can be key to frontier emergence, any effort to preemptively drive institutions (states, financial investors, agribusiness companies…) to focus their investments on sustainable land uses and land use planning efforts can be critical. Learning lessons from the pioneers who have been building and nurturing legacies of previous land use waves can be crucial to avoid repeating mistakes, be they related to local population livelihoods or costly operations failures, which can risk creating large waves of land use expansion with their deforestation and dispossession, for very little subsequent outputs (9,139,140). Steering actors' anticipations towards more sustainable land uses is also possible at this stage. Preemptive or proactive measures can help to increase resilience and adaptability of social-ecological systems to frontier processes as a form of "extreme event" (141). Clarifying and strengthening land tenure rights of local communities, for example, can increase their resilience to sudden frontiers emergence.

Fifth, with this theory, does it mean that we can predict frontier emergence? The short answer is: probably not, as many aspects in this theory remain insufficiently quantified, and contingency and exogenous forces do play a major role in frontier emergence, but we can hopefully identify places that have a certain number of the characteristics that make them *prone* to see frontiers emerge within. Resource availability, exogenous push and subsidy, the build-up of legacies from previous waves, and the rise of actors' anticipations of changes in land rent are the key elements that we identified as decisive to turn a region into a frontier. They can all be explored and monitored to some extent, although exogenous pushes are hard to forecast, and capturing legacies and anticipations require grounded knowledge of the contexts and actors. Inventories, mappings, and forecasts and modeling exercises aiming to identify regions that are prone or vulnerable to sudden large-scale land-use expansion might aim for improved incorporation of such bottom-up knowledge on past land use waves, legacies, and actors' anticipations and expectations, while also investigating whether certain data can constitute adequate proxies for these factors.

# 6. Conclusion

We here synthesized a broad but theoretically-grounded definition of resource and land-use frontiers, as well as different conceptualizations and theories explaining such frontiers. This work is rooted in the context of agricultural, forestry and deforestation frontiers in the Tropics, but can be inspirational for other kinds of frontier processes involving extractive resources, like minerals, energy or conservation frontiers, or even broader forms of "resources" such as opportunities for market, political or religious expansion. From that, we proposed a synthetic theory of emerging frontiers. This theory identifies exogenous pushes, legacies of past waves, and actors' anticipations as the key mechanisms by which frontiers emerged from territories considered marginal in terms of agricultural productivity and connections to global markets, and processes of abnormal rent creation and capture and the progressive built-up of agglomeration economies as the key mechanisms that sustain active frontiers roll out.

We finally discuss five insights derived from our synthesis, on the crucial role of land use policies to steer intensification in resource frontiers contexts; the importance of endogenous dynamics on the sustainability outcomes in frontiers; the pivotal role of emerging frontiers;

and prospects for identifying potential frontier spaces. These insights remain general but might inspire more context-specific works on frontier governance. Together, these insights might help to better understand how to balance development, sustainability and equity concerns in regions that hold abundant resources but might be prone to rapid, uncontrolled, environmentally damaging and inequitable frontier processes.

# References

1. IPBES. The IPBES assessment report on land degradation and restoration. Bonn, Germany: Intergovernmental Science-Policy Platform on Biodiversity and Ecosystem Services; 2018.
2. IPBES. Global Assessment Report on Biodiversity and Ecosystem Services [Internet]. Bonn, Germany: Intergovernmental Science-Policy Platform on Biodiversity and Ecosystem Services; 2019. Available from: http://ipbes.net/global-assessment
3. Castella JC, Lu J, Friis C, Bruun TB, Cole R, Junquera V, et al. Beyond the boom-bust cycle: An interdisciplinary framework for analysing crop booms. Glob Environ Change. 2023;80:102651.
4. Junquera V, Grêt-Regamey A. Crop booms at the forest frontier: Triggers, reinforcing dynamics, and the diffusion of knowledge and norms. Glob Environ Change. 2019;57:101929.
5. Meyfroidt P. Emerging agricultural expansion in northern regions: Insights from land-use research. One Earth. 2021;4(12):1661–4.
6. Price MJ. Seeing Green: Lifecycles of an Arctic Agricultural Frontier. Rural Sociol. 2023;
7. Li TM. Land’s end: Capitalist relations on an indigenous frontier. Duke University Press; 2014.
8. Russo Lopes G, Bastos Lima MG, Reis TNP dos. Maldevelopment revisited: Inclusiveness and social impacts of soy expansion over Brazil’s Cerrado in Matopiba. World Dev. 2021 Mar 1;139:105316.
9. Pendrill F, Gardner TA, Meyfroidt P, Persson UM, Adams J, Azevedo T, et al. Disentangling the numbers behind agriculture-driven tropical deforestation. Science. 2022;377(6611):eabm9267.
10. Lark TJ, Larson B, Schelly I, Batish S, Gibbs HK. Accelerated conversion of native prairie to cropland in Minnesota. Environ Conserv. 2019;46(2):155–62.
11. Pacheco P, Mo K, Dudley N, Shapiro A, Aguilar-Amuchastegui N, Ling PY, et al. Deforestation fronts: Drivers and responses in a changing world. WWF Gland Switz. 2021;125.
12. Buchadas A, Baumann M, Meyfroidt P, Kuemmerle T. Uncovering major types of deforestation frontiers across the world’s tropical dry woodlands. Nat Sustain. 2022;5(7):619–27.
13. Food and Agriculture Organization (FAO). FAOSTAT Statistical database.
14. Baumann M, Gasparri I, Buchadas A, Oeser J, Meyfroidt P, Levers C, et al. Frontier metrics for a process-based understanding of deforestation dynamics. Environ Res Lett. 2022 Sep;17(9):095010.
15. Brunelle T, Makowski D. Assessing whether the best land is cultivated first: A quantile analysis. Plos One. 2020;15(12):e0242222.
16. Eigenbrod F, Beckmann M, Dunnett S, Graham L, Holland RA, Meyfroidt P, et al. Identifying agricultural frontiers for modeling global cropland expansion. One Earth. 2020;3(4):504–14.
17. Kronenburg García A, Meyfroidt P, Abeygunawardane D, Sitoe AA. Waves and

legacies: The making of an investment frontier in Niassa, Mozambique. Ecol Soc. 2022;27(1 (Article No.:) 40).
18. Turner FJ. The significance of the frontier in American history [Internet]. New York: Henry Holt and Company; 1893. Available from: https://www.gutenberg.org/ebooks/22994
19. Barbier EB. Scarcity, frontiers and development. Geogr J. 2012;178(2):110–22.
20. DeFries RS, Foley JA, Asner GP. Land-use choices: balancing human needs and ecosystem function. Front Ecol Environ. 2004;2(5):249–57.
21. Foley JA, DeFries R, Asner GP, Barford C, Bonan G, Carpenter SR, et al. Global Consequences of Land Use. Science. 2005;309(5734):570–4.
22. Hurtt GC, Frolking S, Fearon MG, Moore B, Shevliakova E, Malyshev S, et al. The underpinnings of land-use history: Three centuries of global gridded land-use transitions, wood-harvest activity, and resulting secondary lands. Glob Change Biol. 2006;12(7):1208–29.
23. Ellis EC. Land use and ecological change: A 12,000-year history. Annu Rev Environ Resour. 2021;46:1–33.
24. Fletcher MS, Hamilton R, Dressler W, Palmer L. Indigenous knowledge and the shackles of wilderness. Proc Natl Acad Sci. 2021;118(40):e2022218118.
25. Meyfroidt P, Bremond A de, Ryan CM, Archer E, Aspinall R, Chhabra A, et al. Ten facts about land systems for sustainability. Proc Natl Acad Sci. 2022 Feb 15;119(7).
26. Meyfroidt P, Carlson KM, Fagan ME, Gutiérrez-Vélez VH, Macedo MN, Curran LM, et al. Multiple pathways of commodity crop expansion in tropical forest landscapes. Environ Res Lett. 2014;9(7):074012.
27. Geiger D. Turner in the tropics: The frontier concept revisited [PhD Thesis]. Universität Luzern; 2009.
28. Godfrey BJ, Browder JO. Disarticulated urbanization in the Brazilian Amazon. Geogr Rev. 1996;86(3):441–5.
29. Pacheco P. Populist and capitalist frontiers in the Amazon: Diverging dynamics of a agrarian and land-use change [PhD Thesis]. Clark University; 2005.
30. Southgate DD, Runge CF. The institutional origins of deforestation in Latin America. Staff Pap P USA. 1990;
31. Browder JO, Pedlowski MA, Walker R, Wynne RH, Summers PM, Abad A, et al. Revisiting theories of frontier expansion in the Brazilian Amazon: A survey of the colonist farming population in Rondônia's post-frontier, 1992–2002. World Dev. 2008;36(8):1469–92.
32. Ioris AA. Rent of agribusiness in the Amazon: A case study from Mato Grosso. Land Use Policy. 2016;59:456–66.
33. Fearnside PM. Transmigration in Indonesia: Lessons from its environmental and social impacts. Environ Manage. 1997;21(4):553–70.
34. Moore JW. Sugar and the expansion of the early modern world-economy: Commodity frontiers, ecological transformation, and industrialization. Rev Fernand Braudel Cent. 2000;409–33.
35. Rudel TK. Changing agents of deforestation: from state-initiated to enterprise driven processes, 1970–2000. Land Use Policy. 2007;24(1):35–41.
36. Hecht SB. Soybeans, development and conservation on the Amazon frontier. Dev Change. 2005;36(2):375–404.
37. Brannstrom C. South America's neoliberal agricultural frontiers: places of environmental sacrifice or conservation opportunity? Ambio. 2009;141–9.
38. le Polain de Waroux Y, Baumann M, Gasparri NI, Gavier-Pizarro G, Godar J, Kuemmerle T, et al. Rents, Actors, and the Expansion of Commodity Frontiers in the Gran Chaco. Ann Am Assoc Geogr. 2018 Jan 2;108(1):204–25.
39. Andersson R, Östlund L, Törnlund E. The last European landscape to be colonised: a case study of land-use change in the far north of Sweden 1850-1930. Environ Hist.

2005;11(3):293–318.
40. Merry F, Amacher G, Nepstad D, Lima E, Lefebvre P, Bauch S. Industrial development on logging frontiers in the Brazilian Amazon. Int J Sustain Dev. 2006;9(3):277–96.
41. Ahrends A, Burgess ND, Milledge SAH, Bulling MT, Fisher B, Smart JCR, et al. Predictable waves of sequential forest degradation and biodiversity loss spreading from an African city. Proc Natl Acad Sci. 2010 Aug 17;107(33):14556–61.
42. Klinger JM. Rare earth frontiers: From terrestrial subsoils to lunar landscapes. Cornell University Press; 2018.
43. Orta-Martínez M, Finer M. Oil frontiers and indigenous resistance in the Peruvian Amazon. Ecol Econ. 2010;70(2):207–18.
44. Lord A. Citizens of a hydropower nation: Territory and agency at the frontiers of hydropower development in Nepal. Econ Anthropol. 2016;3(1):145–60.
45. Avila Calero S. Decarbonizing the South. Space, justice and politics at the renewable energy frontiers [Ph.D. Thesis]. Universitat Autònoma de Barcelona; 2021.
46. Guyot S. The eco-frontier paradigm: rethinking the links between space, nature and politics. Geopolitics. 2011;16(3):675–706.
47. Schmink M, Hoelle J, Gomes CVA, Thaler GM. From contested to 'green' frontiers in the Amazon? A long-term analysis of São Félix do Xingu, Brazil. J Peasant Stud. 2019 Feb 23;46(2):377–99.
48. Alvarado L. Institutional Change on a Conservationist Frontier: Local Responses to a Grabbing Process in the Name of Environmental Protection. Land. 2019 Dec;8(12):182.
49. Buchadas A, Qin S, Meyfroidt P, Kuemmerle T. Conservation frontiers: understanding the geographic expansion of conservation. J Land Use Sci. 2022 Jan 19;0(0):1–14.
50. Freitas F. Conservation Frontier. Front Dev Amaz Riches Risks Resist. 2020;51.
51. Laako H, Kauffer E. Between colonising waters and extracting forest fronts: Entangled eco-frontiers in the Usumacinta River Basin. Polit Geogr. 2022 Jun 1;96:102566.
52. Godar J, Gardner T. Governing for sustainability in agricultural-forest frontiers: A case study of the Brazilian Amazon. SEI Discuss Brief. 2014;
53. Schielein J, Börner J. Recent transformations of land-use and land-cover dynamics across different deforestation frontiers in the Brazilian Amazon. Land Use Policy. 2018;76:81–94.
54. Klingler M, Mack P. Post-frontier governance up in smoke? Free-for-all frontier imaginations encourage illegal deforestation and appropriation of public lands in the Brazilian Amazon. J Land Use Sci. 2020;15(2–3):424–38.
55. Meyfroidt P, Lambin EF. Global forest transition: prospects for an end to deforestation. Annu Rev Environ Resour. 2011;36(Journal Article):343–71.
56. Levers C, Romero-Muñoz A, Baumann M, Marzo TD, Fernández PD, Gasparri NI, et al. Agricultural expansion and the ecological marginalization of forest-dependent people. Proc Natl Acad Sci. 2021 Nov 2;118(44).
57. Meyfroidt P, Roy Chowdhury R, de Bremond A, Ellis EC, Erb KH, Filatova T, et al. Middle-range theories of land system change. Glob Environ Change. 2018 Nov 1;53:52–67.
58. Febvre L. Frontière. Bull Cent Int Synthèse. 1928;31–44.
59. Fold N, Hirsch P. Re-thinking frontiers in Southeast Asia. Geogr J. 2009;95–7.
60. Imamura M. Rethinking frontier and frontier studies. Polit Geogr. 2015;(45):96–7.
61. Barbier EB. Scarcity and frontiers: how economies have developed through natural resource exploitation. Cambridge University Press; 2010.
62. Rasmussen MB, Lund C. Reconfiguring Frontier Spaces: The territorialization of resource control. World Dev. 2018;101:388–99.
63. Parker BJ, Rodseth L. Untaming the frontier in anthropology, archaeology, and history. University of Arizona Press; 2005.
64. Tsing AL. Friction: An ethnography of global connection. Princeton University Press; 2011.

65. Altenbernd E, Trimble Young A. Introduction: The significance of the frontier in an age of transnational history. Settl Colon Stud. 2014;4(2):127–50.
66. Smith N. The new urban frontier: Gentrification and the revanchist city. Psychology press; 1996.
67. Lambin EF, Gibbs HK, Ferreira L, Grau R, Mayaux P, Meyfroidt P, et al. Estimating the world's potentially available cropland using a bottom-up approach. Glob Environ Change. 2013;23(5):892–901.
68. Barney K. Laos and the making of a 'relational' resource frontier. Geogr J. 2009;175(2):146–59.
69. Hirsch P. Revisiting frontiers as transitional spaces in Thailand. Geogr J. 2009;175(2):124–32.
70. Arima EY, Richards P, Walker R, Caldas MM. Statistical confirmation of indirect land use change in the Brazilian Amazon. Environ Res Lett. 2011;6(2):024010.
71. Richards PD, Walker RT, Arima EY. Spatially complex land change: The Indirect effect of Brazil's agricultural sector on land use in Amazonia. Glob Environ Change. 2014;29:1–9.
72. Kröger M, Nygren A. Shifting frontier dynamics in Latin America. J Agrar Change. 2020;20(3):364–86.
73. De Jong EB, Knippenberg L, Bakker L. New frontiers: an enriched perspective on extraction frontiers in Indonesia. Crit Asian Stud. 2017;49(3):330–48.
74. Trottier J, Perrier J. Water driven Palestinian agricultural frontiers: The global ramifications of transforming local irrigation. J Polit Ecol. 2018;25(1):292.
75. Kopytoff I. The African frontier: The reproduction of traditional African societies. 1987;
76. Casetti E, Gauthier HL. A formalization and test of the "hollow frontier" hypothesis. Econ Geogr. 1977;53(1):70–8.
77. Rudel TK, Bates D, Machinguiashi R. A tropical forest transition? Agricultural change, out-migration, and secondary forests in the Ecuadorian Amazon. Ann Assoc Am Geogr. 2002;92(1):87–102.
78. Perz SG. Household demographic factors as life cycle determinants of land use in the Amazon. Popul Res Policy Rev. 2001;20:159–86.
79. Walker R, Perz S, Caldas M, Silva LGT. Land use and land cover change in forest frontiers: The role of household life cycles. Int Reg Sci Rev. 2002;25(2):169–99.
80. Coomes OT, Grimard F, Burt GJ. Tropical forests and shifting cultivation: secondary forest fallow dynamics among traditional farmers of the Peruvian Amazon. Ecol Econ. 2000 Jan 1;32(1):109–24.
81. López-Carr D. A review of small farmer land use and deforestation in tropical forest frontiers: implications for conservation and sustainable livelihoods. Land. 2021;10(11):1113.
82. Walker R, Browder J, Arima E, Simmons C, Pereira R, Caldas M, et al. Ranching and the new global range: Amazônia in the 21st century. Geoforum. 2009;40(5):732–45.
83. le Polain de Waroux Y. Capital has no homeland: The formation of transnational producer cohorts in South America's commodity frontiers. Geoforum. 2019;105:131–44.
84. Le Polain de Waroux Y, Garrett RD, Heilmayr R, Lambin EF. Land-use policies and corporate investments in agriculture in the Gran Chaco and Chiquitano. Proc Natl Acad Sci. 2016;113(15):4021–6.
85. Walker R. Theorizing land-cover and land-use change: the case of tropical deforestation. Int Reg Sci Rev. 2004;27(3):247–70.
86. Angelsen A. Policies for reduced deforestation and their impact on agricultural production. Proc Natl Acad Sci. 2010;107(46):19639–44.
87. Lambin EF. Global land availability: Malthus versus Ricardo. Glob Food Secur. 2012;1(2):83–7.
88. Arima EY, Richards P, Walker RT. Biofuel expansion and the spatial economy: implications for the Amazon Basin in the 21st century. Bioenergy Land Use Change.

2017;53–62.
89. Di Tella G. The economics of the frontier. Econ Long View Essays Honour WW Rostow. 1982;210–27.
90. Hecht S, Cockburn A. The fate of the forest: developers, destroyers, and defenders of the Amazon. New York: Harper Perennial; 1990.
91. Schierhorn F, Meyfroidt P, Kastner T, Kuemmerle T, Prishchepov AV, Müller D. The dynamics of beef trade between Brazil and Russia and their environmental implications. Glob Food Secur. 2016 Dec 1;11:84–92.
92. Hall D. Land grabs, land control, and Southeast Asian crop booms. J Peasant Stud. 2011;38(4):837–57.
93. Porter ME. Competitive advantage, agglomeration economies, and regional policy. Int Reg Sci Rev. 1996;19(1–2):85–90.
94. Scott A, Storper M. Regions, globalization, development. Reg Stud. 2003;37(6–7):579–93.
95. Krugman P. Increasing returns and economic geography. J Polit Econ. 1991;99(3):483–99.
96. Porter M. Competitive advantage of nations. New York: The Free Pres; 1990.
97. Abeygunawardane D, Kronenburg García A, Sun Z, Müller D, Sitoe A, Meyfroidt P. Resource frontiers and agglomeration economies: The varied logics of transnational land-based investing in Southern and Eastern Africa. Ambio. 2022;51(6):1535–51.
98. Garrett RD, Lambin EF, Naylor RL. The new economic geography of land use change: supply chain configurations and land use in the Brazilian Amazon. Land Use Policy. 2013;34:265–75.
99. Richards P. It’s not just where you farm; it’s whether your neighbor does too. How agglomeration economies are shaping new agricultural landscapes. J Econ Geogr. 2018;18(1):87–110.
100. Glassman J. Primitive accumulation, accumulation by dispossession, accumulation by ‘extra-economic’ means. Prog Hum Geogr. 2006;30(5):608–25.
101. Harvey D. The new imperialism. Oxford, UK: Oxford University Press; 2005.
102. Peluso NL, Lund C. New frontiers of land control: Introduction. J Peasant Stud. 2011;38(4):667–81.
103. Kronenburg García A, Dijk H van. Towards a Theory of Claim Making: Bridging Access and Property Theory. Soc Nat Resour. 2020 Feb 1;33(2):167–83.
104. Watts MJ. Frontiers: Authority, Precarity and Insurgency at the Edge of the State. World Dev. 2018 Jan 1;(101):477–88.
105. Lima MGB, Kmoch L. Neglect paves the way for dispossession: The politics of “last frontiers” in Brazil and Myanmar. World Dev. 2021;148:105681.
106. Vandergeest P, Peluso NL. Territorialization and state power in Thailand. Theory Soc. 1995;385–426.
107. Korf B, Hagmann T, Emmenegger R. Re-spacing African drylands: Territorialization, sedentarization and indigenous commodification in the Ethiopian pastoral frontier. J Peasant Stud. 2015;42(5):881–901.
108. Olofsson M. Expanding commodity frontiers and the emergence of customary land markets: A case study of tree-crop farming in Venda, South Africa. Land Use Policy. 2021;101:105203.
109. De Koninck R. The theory and practice of frontier development: Vietnam’s contribution. Asia Pac Viewp. 2000;41(1):7–21.
110. Meyfroidt P, Vu TP, Hoang VA. Trajectories of deforestation, coffee expansion and displacement of shifting cultivation in the Central Highlands of Vietnam. Glob Environ Change. 2013;23(5):1187–98.
111. Nyerges AE. The Ecology of Wealth-in-People: Agriculture, Settlement, and Society on the Perpetual Frontier. Am Anthropol. 1992;94(4):860–81.
112. Korf B, Raeymaekers T. Introduction: Border, Frontier and the Geography of Rule at the

Margins of the State. Violence Margins States Confl Borderl. 2013;3–27.
113. Yiftachel O. The internal frontier: territorial control and ethnic relations in Israel. Reg Stud. 1996;30(5):493–508.
114. Martinez-Alier J, Temper L, Del Bene D, Scheidel A. Is there a global environmental justice movement? J Peasant Stud. 2016 May 3;43(3):731–55.
115. Zoomers A. Introduction: Rushing for Land: Equitable and sustainable development in Africa, Asia and Latin America. Development. 2011 Mar 1;54(1):12–20.
116. Dietz K, Engels B. Analysing land conflicts in times of global crises. Geoforum. 2020;111:208–17.
117. Wolford W. The colonial roots of agricultural modernization in Mozambique: the role of research from Portugal to ProSavana. J Peasant Stud. 2021 Feb 23;48(2):254–73.
118. Mosca J. Políticas públicas e agricultura em Moçambique. Escolar Editora; 2016.
119. Weszkalnys G. Anticipating oil: the temporal politics of a disaster yet to come. Sociol Rev. 2014;62(S1):211–35.
120. Kronenburg García A, Meyfroidt P. Pioneers, emerging frontiers and economies of anticipation in northern Mozambique. Prep.
121. Tsing A. Inside the Economy of Appearances. Public Cult. 2000 Jan 1;12(1):115–44.
122. Junquera V, Schlüter M, Rocha J, Wunderling N, Levin SA, Rubenstein DI, et al. Crop booms as regime shifts. Revis. 2024;
123. Ioris AA. Amazon’s dead ends: Frontier-making the centre. Polit Geogr. 2018;65:98–106.
124. Rodrigues ASL, Ewers RM, Parry L, Souza C, Veríssimo A, Balmford A. Boom-and-Bust Development Patterns Across the Amazon Deforestation Frontier. Science. 2009 Jun 12;324(5933):1435–7.
125. Shivaram R, Biggs NB. Frontiers in multi-benefit value stacking for solar development on working lands. Environ Res Lett. 2023 Jan;18(1):011002.
126. Barbier EB. Scarcity, frontiers and the resource curse: A historical perspective 1. In: Natural Resources and Economic Growth. Routledge; 2015.
127. Boserup E. The Conditions of Agricultural Growth: The Economics of Agrarian Change Under Population Pressure. London: Routledge; 1965. 124 p.
128. Turner BL, Ali AMS. Induced intensification: Agricultural change in Bangladesh with implications for Malthus and Boserup. Proc Natl Acad Sci. 1996 Dec 10;93(25):14984–91.
129. Rodriguez Garcia V, Gaspart F, Kastner T, Meyfroidt P. Agricultural intensification and land use change: Assessing country-level induced intensification, land sparing and rebound effect. Environ Res Lett [Internet]. 2020; Available from: http://hdl.handle.net/2078.1/230899
130. Garrett RD, Koh I, Lambin EF, le Polain de Waroux Y, Kastens JH, Brown JC. Intensification in agriculture-forest frontiers: Land use responses to development and conservation policies in Brazil. Glob Environ Change. 2018 Nov 1;53:233–43.
131. Koch N, zu Ermgassen EK, Wehkamp J, Oliveira Filho FJ, Schwerhoff G. Agricultural productivity and forest conservation: evidence from the Brazilian Amazon. Am J Agric Econ. 2019;101(3):919–40.
132. Moffette F, Skidmore M, Gibbs HK. Environmental policies that shape productivity: Evidence from cattle ranching in the Amazon. J Environ Econ Manag. 2021 Sep 1;109:102490.
133. Marin FR, Zanon AJ, Monzon JP, Andrade JF, Silva EH, Richter GL, et al. Protecting the Amazon forest and reducing global warming via agricultural intensification. Nat Sustain. 2022;5(12):1018–26.
134. Jepson W. Producing a Modern Agricultural Frontier: Firms and Cooperatives in Eastern Mato Grosso, Brazil. Econ Geogr. 2006;82(3):289–316.
135. Garrett RD, Lambin EF, Naylor RL. Land institutions and supply chain configurations as determinants of soybean planted area and yields in Brazil. Land Use Policy.

2013;31:385–96.
136. VanWey LK, Spera S, de Sa R, Mahr D, Mustard JF. Socioeconomic development and agricultural intensification in Mato Grosso. Philos Trans R Soc B Biol Sci. 2013 Jun 5;368(1619):20120168.
137. Oliveira E, Meyfroidt P. Strategic land-use planning instruments in tropical regions: state of the art and future research. J Land Use Sci. 2021;16(5–6):479–97.
138. Oliveira E, Meyfroidt P. Strategic spatial planning in emerging land-use frontiers: evidence from Mozambique. Ecol Soc. 2022;27(2).
139. Borras Jr SM, Franco JC, Moreda T, Xu Y, Bruna N, Demena BA. The value of so-called 'failed' large-scale land acquisitions. Land Use Policy. 2022;119:106199.
140. Buchadas A, Meyfroidt P, Baumann M, Lu J, Kronenburg García A, Mastrangelo M, et al. Unpacking the role of failed land investment projects in enabling and driving deforestation. Submitt Manuscr.
141. Levin SA, Anderies JM, Adger N, Barrett S, Bennett EM, Cardenas JC, et al. Governance in the face of extreme events: Lessons from evolutionary processes for structuring interventions, and the need to go beyond. Ecosystems. 2021;1–15.